\documentclass[twoside,10.5pt]{article}%                        *
\usepackage{mathrsfs}% *
\usepackage{float}

\usepackage{multirow}
\usepackage{pifont}%  
\usepackage{longtable}
\usepackage{amsmath}% 
\usepackage{hyperref}
\usepackage{booktabs}
\usepackage{tabularx}
\usepackage{amsthm}%  
\usepackage{endnotes}
\usepackage{txfonts}%                                           *
 \usepackage{geometry}%                                         *
\usepackage{latexsym}%                                          *
\usepackage{amssymb}%                                           *
\usepackage{graphicx}%                                          *
\usepackage{geometry}%                                          * Please do not change any words here.
\usepackage{ifthen}%                                            *
\usepackage{fancyhdr}% 
\usepackage{comment}
\usepackage{afterpage}%                                         *
\usepackage{xcolor}%                                            *
\usepackage{ifthen}%                                            *
\usepackage{titlesec}%                                          *
\titleformat{\section}{\bfseries}{\thesection.}{0.3em}{}%       *
\titlespacing{\section}{0pt}{0pt}{0pt}%                         *
\titleformat{\subsection}{\itshape}{\thesubsection}{0.3em}{}%   *
\titlespacing{\subsection}{0pt}{0pt}{0pt}%                      *
\titleformat{\subsubsection}{\rmfamily}%                        *
{\thesubsubsection}{0.3em}{}%                                   *
\titlespacing{\subsubsection}{0pt}{0pt}{0pt}%                   *
\let\footnote=\endnote

\begin{document}
%================TITLE===========================================
\begin{center}
{\LARGE{A
Formalization of Group Decision Making in Multi-viewpoints Design}}\\[15pt]
\end{center}
%====================================================
\begin{center}
Saloua Bennani$^{1,2}$, Iliass Ait El Kouch$^{3}$, Mahmoud El Hamlaoui$^1$, Sophie Ebersold$^2$, Bernard Coulette$^2$ \& Mahmoud Nassar$^1$
\end{center}
$^1$ IMS Team, ADMIR Laboratory, ENSIAS, Rabat IT Center, Mohammed V University in Rabat,
Morocco \par
$^2$ SMART Team, IRIT Laboratory, University of Toulouse Jean Jaures (UT2J), Toulouse, France \par
$^3$ ENSIAS, Rabat, Morocco \par
Correspondence: Saloua Bennani, IMS Team, ADMIR Laboratory, ENSIAS, Rabat IT Center, Mohammed V
University in Rabat, Morocco. Tel: 212-648-814-157. E-mail: saloua.bennani91@gmail.com\\[15pt]
%================ABSTRACT====================================

\textbf{Abstract}

Complex systems are typically designed collaboratively by stakeholders from different domains. This multi viewpoints paradigm promotes the separation of concerns since separate teams, from different business viewpoints, build partial models describing the system. These partial models are naturally heterogeneous. So, it is difficult to ensure their inter-model consistency if kept separately. For that, we propose a collaborative approach that combines Group Decision Making (GDM) and Model-Based Engineering (MBE). This paper highlights the GDM part of our approach and especially the concept of decision policy that enables coming up with collective decisions in group decision-making contexts.  

%================KEYWORDS====================================
\textbf{Keywords:} group decision-making (GDM), collaboration, pattern, design, multi viewpoints, models

%================MAIN TEXT====================================
\textbf{1. Introduction}

Complex systems are typically designed collaboratively by stakeholders from different domains. Each of these domains depicts a given view on the system (e.g., physical or software view). Thus, the design of the system as a whole requires a multi-view modeling approach in which the complexity of the system is reduced since each team focuses on a given point of view (i.e., separation of concerns principle (Aksit, 1996; France et al., 2003)).

Mechatronic systems are good examples of complex systems that involve, in their design, teams from various disciplines. Designing such a system will produce a set of partial models dealing with mechanical, electrical and computing viewpoints, etc. The main issue with this separation of concerns is the way we can ensure, for a given system, the consistency among its partial models. Furthermore, these models may evolve. So, managing them separately might cause the global system inconsistency. Some research works attempt to solve this issue by relating the viewpoints, this is called \textit{model matching}, \textit{inter-model relationships}, or even \textit{multi-view consistency} (Nuseibeh et al., 2000; Cicchetti et al., 2019; Feldmann et al., 2019).

We put ourselves in a model based engineering context and analyze works done in this field. One of the following two limitations apply to most of the existing  approaches : 
\begin{itemize}
\item The approach provides a fixed set of relationships to relate models, which could jeopardize the completeness of the produced correspondences among partial models (Zhdanova et Shvaiko, 2006; Br{\"a}uer et Lochmann, 2008; Shosha et al., 2015). Note that we consider a correspondence as  a relationship that relates at least two elements.
\item The approach assumes that a unique actor can perform solely the alignment, i.e., define the correspondences needed to relate models of a specific application domain. This could challenge the accuracy and validity of the produced correspondences (Bruneliere et al., 2015; Golra et al., 2016; El Hamlaoui et al., 2018). 
\end{itemize}

We propose a collaborative approach to define inter-model correspondences. This approach is based on two lines of work: model alignment  and group decision-making (GDM). In this paper, we focus on the GDM part of the approach. In Section 2, we summarize the proposed metamodel MMCollab, then we detail the collective decision elaboration through the instantiation of its concept GDMPattern, which we call DecisionPolicy.  The implementation of DecisionPolicy is made thanks to the State and Observer patterns (Gamma, 1995) to comply with design best practices. In Section 3, we apply the proposed GDM formalization to align models of a conference management system. Section 4 presents the related work in terms of GDM modelling and formalisation.  The last section concludes the paper.

\textbf{2. Method}

\emph{2.1 CAHM Approach overview}

\begin{figure*}[htbp!]
\center
{\includegraphics[scale=0.3]{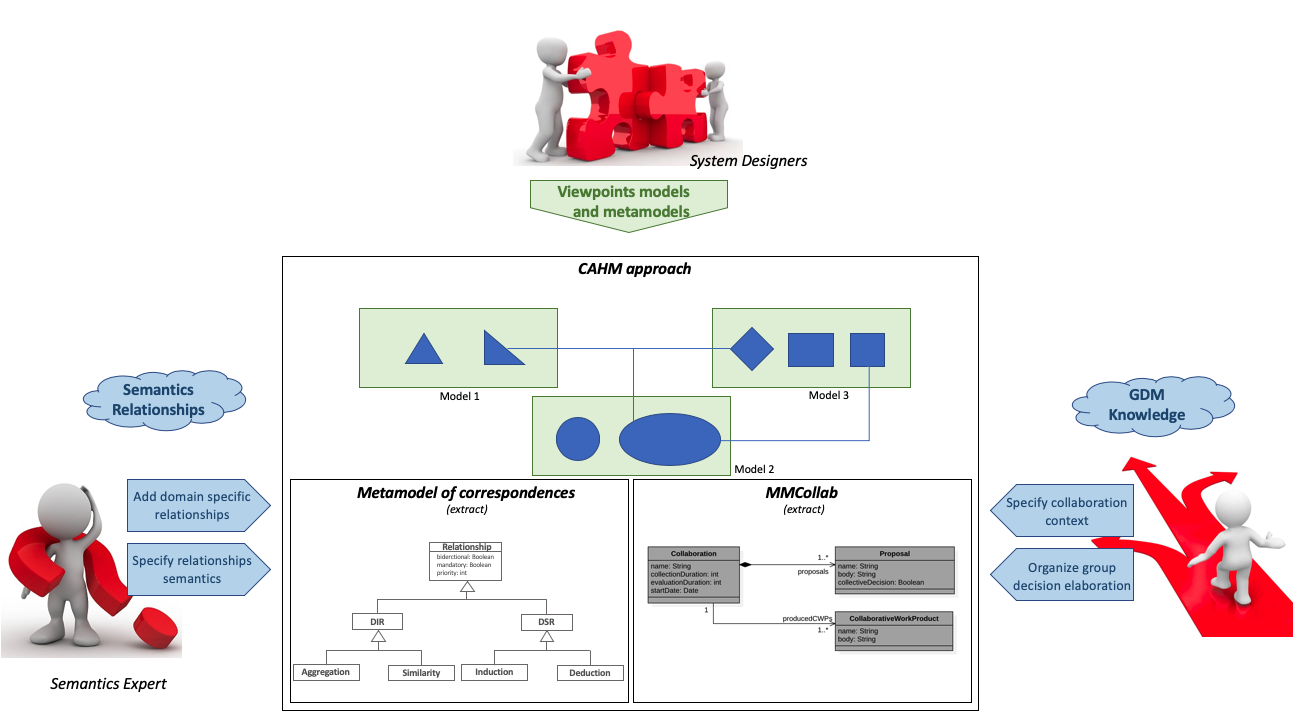}}

\caption{\label{cahm} Overview of the CAHM approach}
\end{figure*}

The approach we propose, CAHM (for \textit{Collaborative Alignment of Heterogeneous Models}), essentially brings together two lines of work: model alignment and GDM. The goal of CAHM is to enhance decision-making in case of heterogeneous models alignment. This approach is based on two main elements as  Figure \ref{cahm} shows: MMCollab and MMC.

MMCollab (See Section 2.2) organizes GDM knowledge by providing concepts and relationships covering proposals elaboration, decision policies definition, actors and GDM enactment. MMCollab's application context is wider than model alignment. In fact, it can be used in other fields where a GDM process is required. 

The second element is the metamodel of correspondences (MMC). Its aim is to carry out relationships and their semantics in order to make them usable within a semi-automatic process for model alignment. By operating the semantics of relationships, the approach reduces the human contribution.

Two groups of users can use this approach. The first one is viewpoints designers, called local coordinators, who provide proposals (meta-correspondences, i.e., correspondences at metamodel level) and evaluate them collaboratively. The second gathers approach experts: (1) semantics experts who define the relationships that may relate models and implement their semantics and (2) GDM experts who define the characteristics of decision policies and make them accessible for use by the local coordinators.

\emph{2.2 MMCollab Overview}

\begin{figure*}[htbp!]
\center
{\includegraphics[scale=0.25]{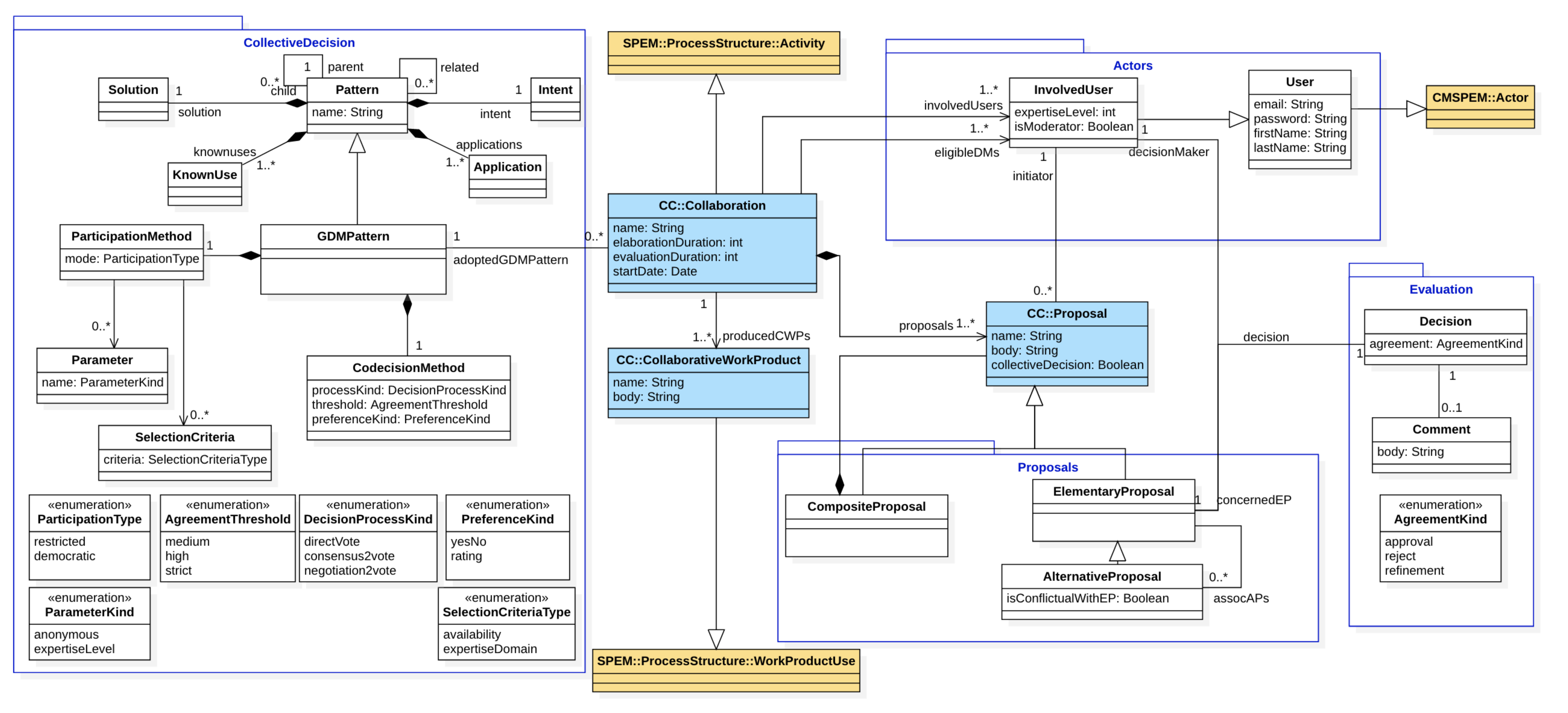}}

\caption{\label{mmcollab} Overall view of metamodel of collaboration (MMCollab)}
\end{figure*}

MMCollab, shown in Figure \ref{mmcollab}, was introduced in (Bennani et al., 2018) and completed in (Bennani et al., 2019). It aims to formalize group decision-making processes. Here, we describe  briefly its concepts since the core of this paper depends on it. Table \ref{tab1} gives a summary of all MMCollab's concepts.

\emph{Collaboration} is the central concept of MMCollab. It specializes the SPEM's Activity\footnote{SPEM: https://www.omg.org/spec/SPEM/2.0/PDF} and contains a set of \textit{Proposal}s. A \emph{Collaboration} is enacted according to a \emph{GDMPattern} and a \emph{collectiveDecision} is attributed to each \emph{Proposal} at the end of the collaborative session.

A \emph{Collaboration} requires a set of \emph{involvedUsers}, including a moderator. This latter has to choose the \emph{GDMPattern} to be followed in the collaboration: \emph{adoptedGDMPattern}.  A list of eligible decision makers (\emph{eligibleDMs}) is initialized by the \emph{involvedUsers} who satisfy the \emph{adoptedGDMPattern}.
A \emph{Proposal} may be composite or elementary. An \emph{InvolvedUser} that can evaluate a \emph{Proposal} is called a \emph{decisionMaker}. His evaluation associates an individual decision (\emph{Decision}) to each proposal.
Three decisions are possible:  \emph{approval}, \emph{reject} or \emph{refinement} (enumeration: \emph{AgreementKind}). In case of a reject, a \emph{Comment} should be given to justify the decision. In case the \emph{decisionMaker} thinks an EP needs to be refined, he/she provides an \emph{AlternativeProposal} (AP). 
The value of \emph{collectiveDecision} attribute of a \emph{Proposal} is determined by combinig the individual \emph{Decisions} according to the \emph{adoptedGDMPattern}. The products of a \emph{Collaboration} are called \emph{CollaborativeWorkProduct(s)}, they gather the approved proposals.

\begin{table}[htbp!]
\centering
\caption{\label{tab1}Summary of MMCollab's concepts}

\begin{tabularx}{\textwidth}{l l l X}

\toprule{}
%\multicolumn{3}{c}{\textbf{Meta-Correspondence}} & \multicolumn{2}{c}{\textbf{Evaluation}} \\
  
\textbf{Package} & \textbf{Concept}  & \textbf{Super-Class}  &
\textbf{Description} \\
\midrule
\multirow{3}{*}{CoreConcepts}& Collaboration & 
 
SPEM:PS:Activity & - consists of the development and evaluation of a set of proposals.  \\

& CollaborativeWorkProduct & 
SPEM:PS:WorkProductUse & - the outputs of a Collaboration.  \\
 & Proposal & 
-- &  - the subject of decision-making.  \\

\midrule

\multirow{2}{*}{Actors}& User & CMSPEM:PS:Actor\footnote{CMSPEM: Kedji, K. A., Lbathd, R., Coulette, B., Nassar, M., Baresse, L., \& Racaru, F. (2012). Supporting collaborative development using process models: An integration-focused approach. In \emph{2012 International Conference on Software and System Process (ICSSP)} (pp. 120-129). IEEE.} & 
- an actor of a Collaboration.  \\
& InvolvedUser & User & 
- specifies the user's role (\textit{isModerator}) and weights (\textit{expertiseLevel}).\\
\midrule

\multirow{3}{*}{Proposals}& CompositeProposal & 
Proposal
 &  - a tree of EP approved or rejected together.   \\

& ElementaryProposal (EP) & 
Proposal
 & - the leaves of the composite pattern (CompositeProposal).  \\
 & AlternativeProposal (AP) & 
ElementaryProposal &  - provided when a decision maker thinks an EP needs to be refined.  \\
\midrule

\multirow{3}{*}{Evaluation}& Comment & 
--
 &  - justifies a given decision. It is mandatory in case of reject.
  \\

& Decision & 
--
 & - the position of an actor on an EP.  \\
 & AgreementKind & 
-- &  
- the possible values of agreement: \textit{approval}, \textit{reject} or \textit{refinement}.

Refinement means that EP have to be adjusted using an AP.
 \\

\midrule

\multirow{7}{*}{CollectiveDecision}& Pattern & 
--
 & - has an \textit{Intent}, a \textit{Solution}, some \textit{Known uses} and \textit{Application contexts}.  
  \\

 & CoDecisionMethod & -- &  - the collective decision-making method that consists of a processKind (\textit{DecisionProcessKind}), a threshold (\textit{AgreementThreshold}) and a preference (\textit{PreferenceKind}). \\
 
& DecisionProcessKind & -- & - defines the type of process: 
\textit{directVote}, \textit{consensus2vote}, \textit{negociation2vote}.\\
& AgreementThreshold & -- & 
- summarizes the potential values of acceptance threshold: \textit{low}, \textit{medium}, \textit{high} and \textit{strict}.\\
&PreferenceKind & -- &- indicates how proposals evaluation is performed : \textit{rating}, or \textit{yesNo}. \\
&ParticipationMethod & -- & - a participation is considered as \textit{democratic} when all involvedUsers participate, otherwise it is \textit{restricted}. 

For restricted participation, the criteria of actors selection should be specified.
 \\
& GDMPattern & Pattern & - a specialization of Pattern that precises how the collective decision is elaborated. It consists of a \textit{CoDecisionMethod} and a \textit{ParticipationMethod}. \\

 \bottomrule 

\end{tabularx}
\end{table}

\emph{2.3 DecisionPolicy from a conceptual point of view}

We consider now instances of a \textit{GDMPattern}, called \textit{DecisionPolicy (DP)}. Actually, a DP is defined by combining instances of elements that characterize a GDMPattern (i.e., \textit{ParticipationMethod} and \textit{CoDecisionMethod}) and by transitivity a combination of instances of elements that characterize both of them (i.e., type of participation (type), decision process (processKind), agreement threshold (threshold) and preference kind (preferenceKind)). 

\begin{figure*}[htbp!]
\center
{\includegraphics[scale=0.5]{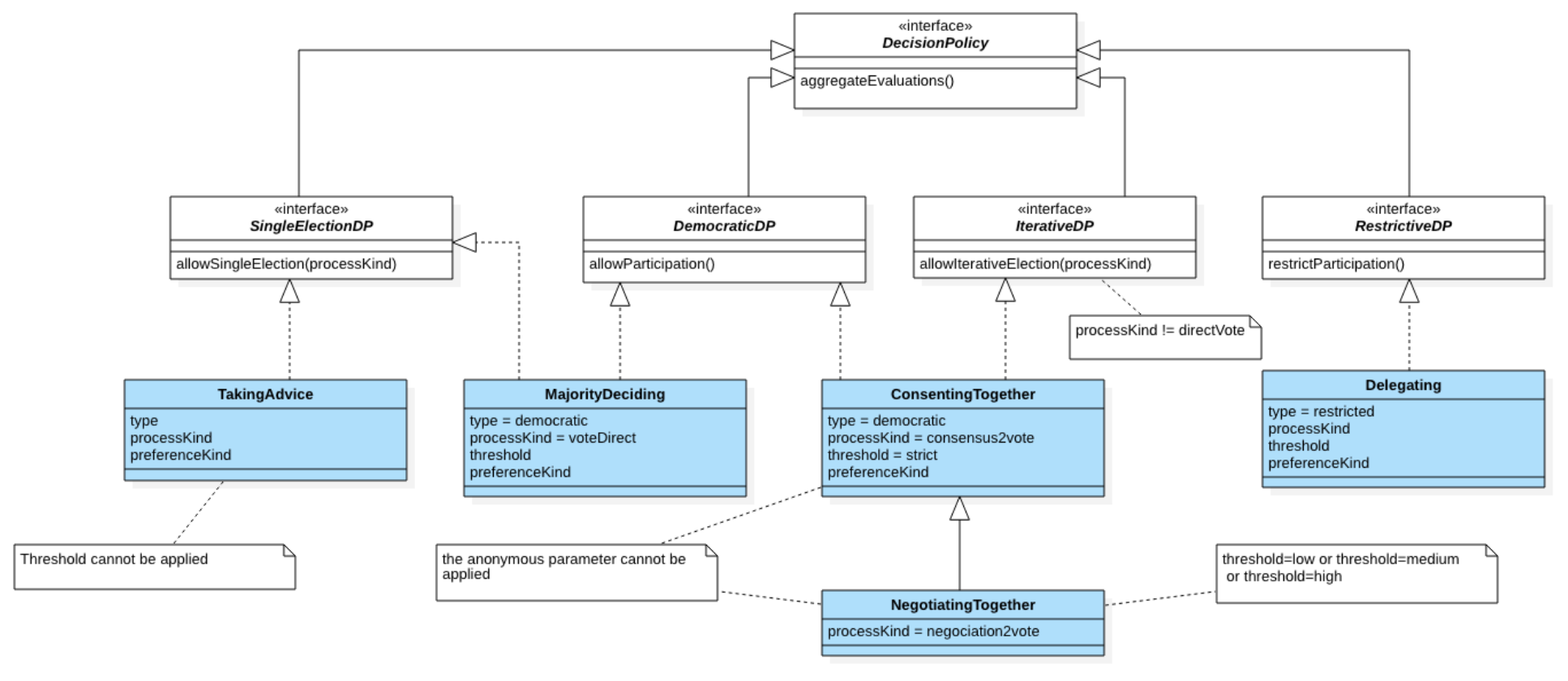}}

\caption{\label{dps} 
Decision policies structuration}
\end{figure*}

Five \textit{decision policies} have been defined by combining these elements. They represent the commonly used policies, namely: \emph{Delegating}, \emph{Taking advice}, \emph{Majority deciding}, \emph{Consenting together} and \emph{Negotiating together} as Figure \ref{dps} shows.

These five decision policies have been organized according to (i) their type of participation: \emph{Restricted (RestrictedDP) vs Democratic (DemocraticDP)}, and (ii) the number of turns needed to come up with a decision: \emph{SingleElectionDP vs IterativeDP}. 

\begin{itemize}
\item \emph{Delegating} and \emph{Taking advice} are restricted decision policies, which means that the criteria for selecting actors must be specified. \emph{Delegating} can delegate the decision taking to a single person or a subset of stakeholders that satisfy the conditions of delegates' choice. \emph{Delegating} can be done in one or multiple rounds. 
\emph{Taking advice} is performed in one turn; the decision maker takes other people's advice and it is up to him alone to make the decision.

\item \emph{Majority deciding} is a democratic decision policy that specializes \emph{SingleElectionDP} since it requires only one turn to be performed. Thus, if the fixed threshold is not reached, either the stakeholders agree with the moderator to adjust the threshold, or they re-evaluate the proposals.

\item \emph{Consenting together} and \emph{Negotiating together} are \emph{IterativeDP}, which means they may be repeated until the fixed threshold is reached.
\emph{Consenting together} requires a \emph{strict} threshold (100\% acceptance) while \emph{Negotiating together} works with a lower value. 
\end{itemize}
These decision policies are not frozen and can be extended according to the requirements of application contexts, by exploring the possible combinations of  elements that define them. For example, the \emph{processKind} and \emph{threshold} for \emph{Delegating} decision policy are not fixed. So, they can take every possible value and provide a decision policy similar to Majority deciding, Consenting together and Negotiating together but in a restrictive mode.

To facilitate choosing among these decision policies, we provide a descriptive manual that represent them following the widespread structuration of patterns and which correspond to the characteristics of the Pattern concept (i.e., \emph{name, intent, applications, solution, known uses, related patterns}).  Table \ref{tab2} describes the Majority deciding policy following this structure.

\begin{table}[H]
\centering
\caption{\label{tab2}Description of majority deciding decision policy}
\begin{tabularx}{\textwidth}{ lX }
\toprule{}
%\multicolumn{3}{c}{\textbf{Meta-Correspondence}} & \multicolumn{2}{c}{\textbf{Evaluation}} \\
  
\textbf{Name} & \emph{Majority Deciding} 

\\
\textbf{Intent} & 
Reach a decision that takes into account the opinions of all the stakeholders. The proposal(s) approved by the majority of the group is (are) adopted.

 \\
\textbf{Applications} & 
This pattern is to be used in case:\\
& - decision makers competencies and weights are almost equal.\\
& - time constraints: it requires less time since it is done in a single turn. 

\\

\textbf{Known uses} & Single-round elections either held in face-to-face or by electronic vote.

\\
\textbf{Solution} & This pattern enactment goes through five steps. First, the moderator defines the collaboration characteristics  (intent and duration). Then, he/she sets the threshold and preferenceKind of the codecision method (the processKind is set to voteDirect). Afterwards, he/she notifies the actors concerned to whom he/she assigns the role decision maker.

If the proposals are not already established, decision makers start by drawing up the list of proposals. Then, they express their individual preferences. At the end, a tool (or possibly the moderator) aggregates individual preferences and proposals exceeding the threshold are approved and constitute the group decision. 
Several proposals can be approved if they are not conflicting.
\begin{center}
 \includegraphics[width=0.75\textwidth, height=40mm]{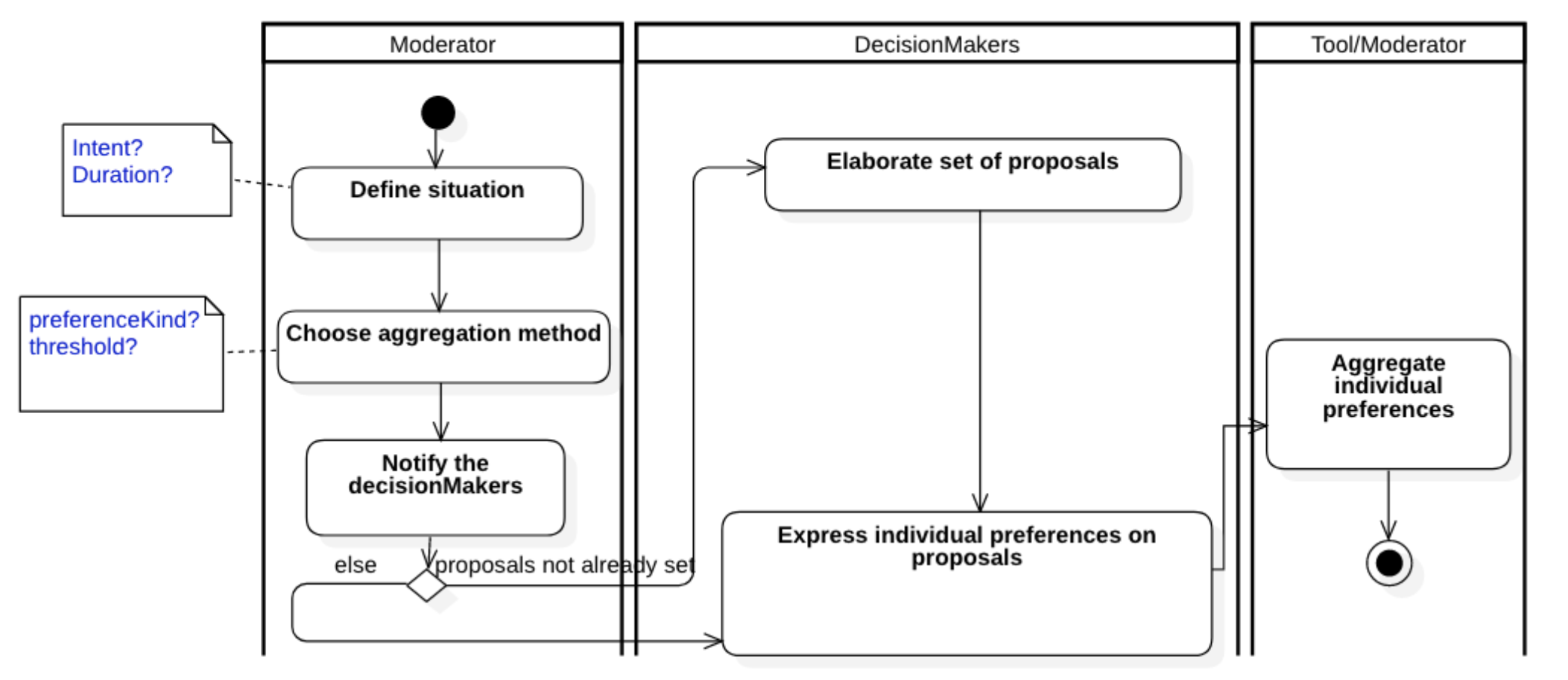}
\end{center}\\

\textbf{Related patterns} & Delegating \\
&  \emph{Majority Deciding} and \emph{Delegating} differ in the type of participation and the  actors' weight. Delegating makes a prior choice of the involved actors while Majority deciding is democratic.\\

\bottomrule
\end{tabularx}
\end{table}

\emph{2.4 DecisionPolicy implementation using design best practices}

To implement the concept DecisionPolicy, two known design patterns have been used: State and Observer (Gamma, 1995). The first one is used to characterize and implement all the states of a decision policy, while the second one is used when there is one to many relationships between objects so that if one object is modified, its dependent objects are to be notified automatically and updated.

2.4.1 Use of State pattern

Decision policies define how individual decisions will be aggregated. There are iterative and single-round strategies. We use the State pattern to allow the DecisionPolicy to alter its behavior when its internal state changes. The State pattern is used in computer programming to encapsulate behaviors of  the same object, based on its internal state. This is a clean way for an object to change its behavior at runtime without resorting to conditional statements.

\begin{figure*}[htbp!]
\center
{\includegraphics[scale=0.48]{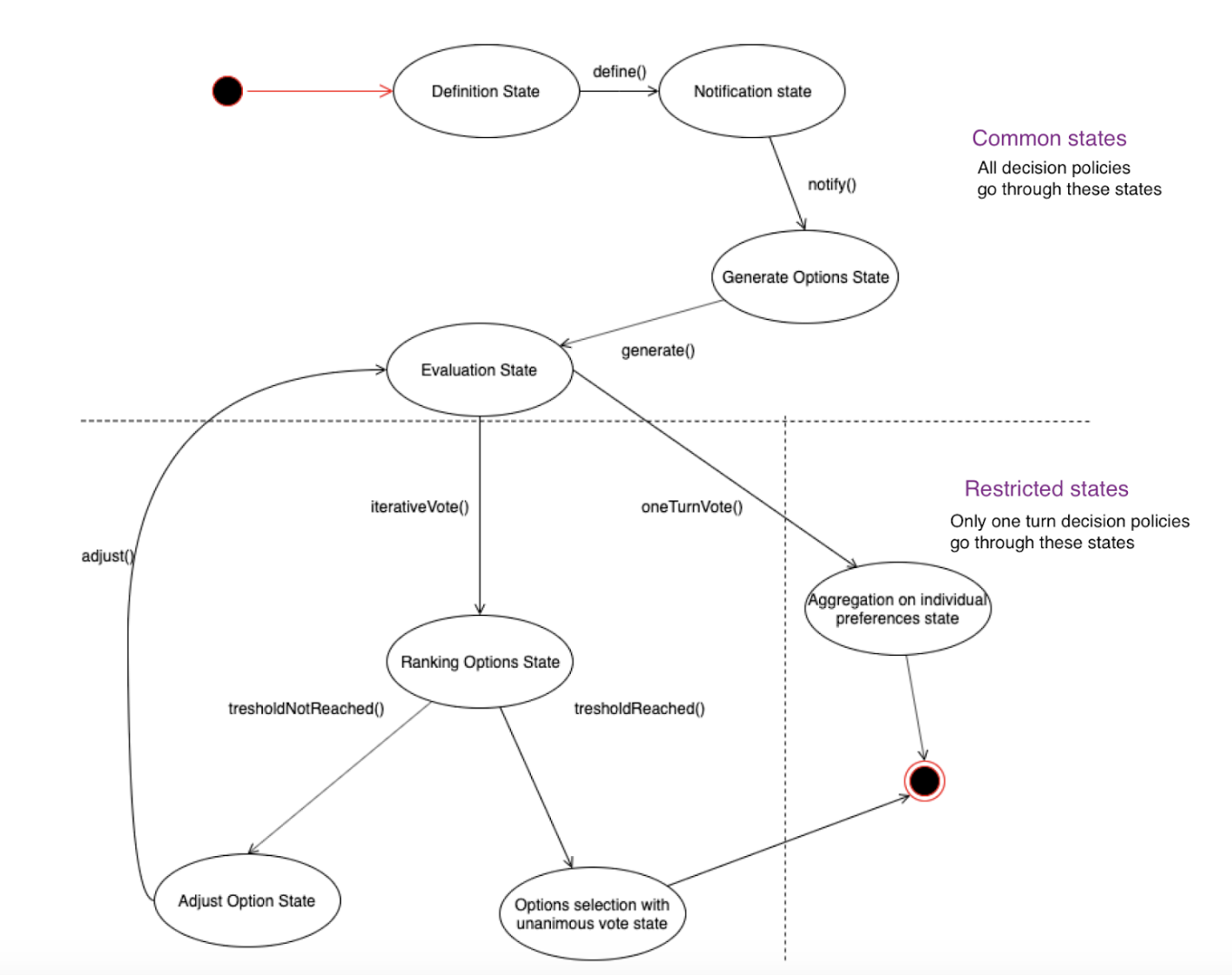}
\caption{\label{stateMachine} State machine of a decision policy}}

\end{figure*}

Whether the decision policy is one-round or iterative, the moderator must define the collaboration situation and choose the aggregation method, then notify the concerned decision makers and let them set the proposals and evaluate them.
In case the decision policy is iterative and the threshold set by the moderator is not reached after the evaluation step (for example, 60\% acceptance reached whereas the threshold set to 80\%), the decision makers have to adjust the proposals until the threshold is  reached.
In the other case (a single-round decision policy), directly after the evaluation step, the tool or the moderator assesses whether the proposals will be approved or not.
Figure \ref{stateMachine} presents a state machine that distinguishes the common states of a decision policy from the restricted ones. Common states are states adapted to all decision policies either they are iterative or single-rounds, whereas restricted states are specific to one turn decision policies.

\begin{figure*}[htbp!]
\center
{\includegraphics[scale=0.5]{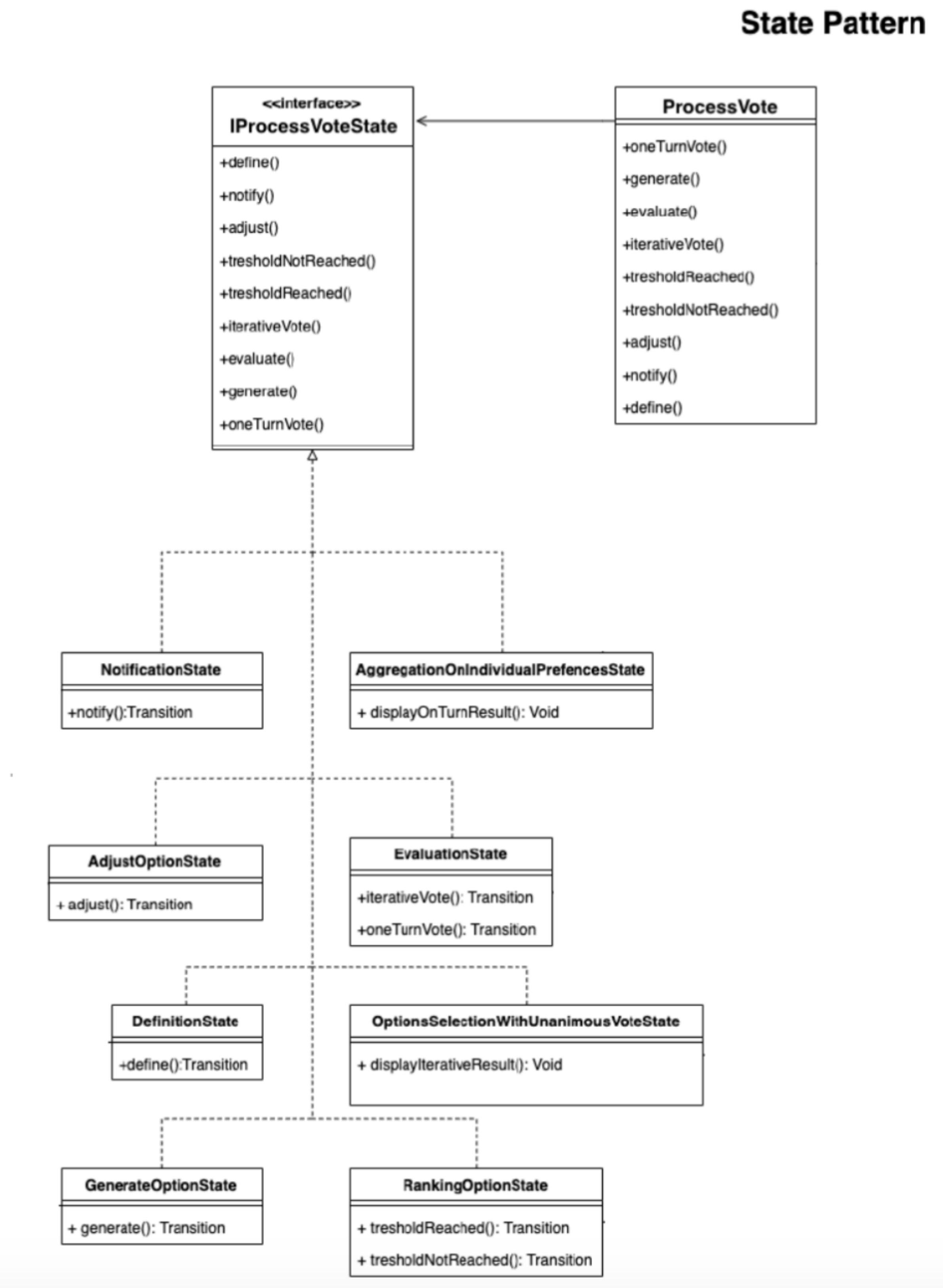}}

\caption{\label{statePattern} State design pattern of a decision policy}
\end{figure*}

Based on the state machine of Figure \ref{stateMachine}, we obtain the class diagram presented in Figure \ref{statePattern}, on which the interface IProcessVote defines the potential actions of all states.

2.4.2 Use of Observer pattern 

\begin{figure*}[htbp!]
\center
{\includegraphics[scale=0.48]{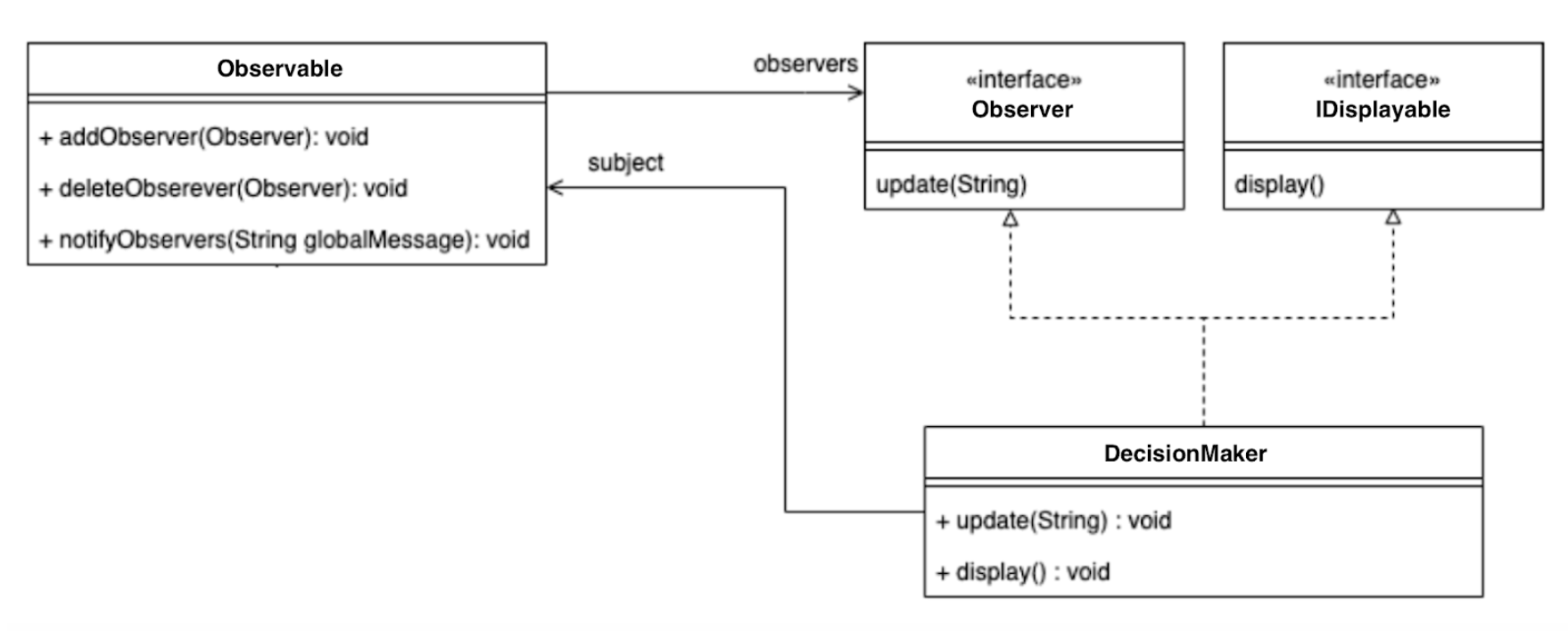}}

\caption{\label{observerPattern}Use of Observer pattern for notifying a proposal's decision makers}
\end{figure*}

Observer is used to select and notify the concerned decision makers for each proposal. In fact, this pattern allows an object, called Observable in Figure \ref{observerPattern}, to maintain a list of its dependents, called observers, and automatically notifies them of any state change, usually by calling one of their methods (the update() method).
The responsibility of observers is to register (and unregister) themselves on Observable (to get notified of state changes) and to update their state (synchronize their state with Observable state) when they are notified.
This makes Observable and observers loosely coupled. Observable and observers have no explicit knowledge of each other. Observers can be added and removed independently at run-time.

\emph{2.5 HMCS-Collab Tool}

\begin{figure*}[htbp!]
\center
{\includegraphics[scale=0.45]{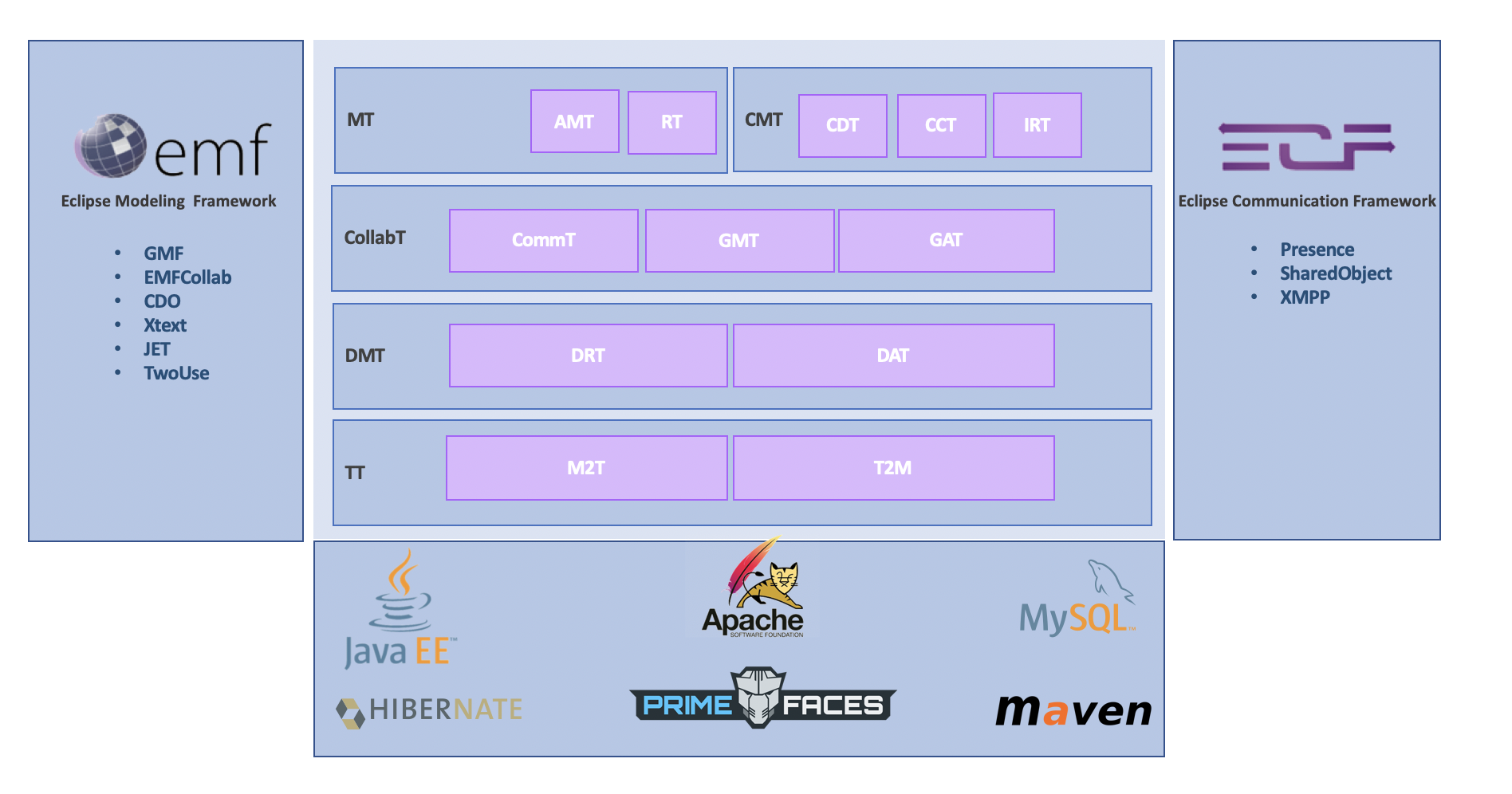}}

\caption{\label{architecture} Architecture of HMCS-Collab
}
\end{figure*}

To support collaborative model alignment, we have developed a support tool called HMCS-Collab. It is based on HMCS Tool for model matching (we use its modules MT, CMT and TT and enhance the definition of semantics relationships). Figure \ref{architecture} presents the global architecture of HMCS-Collab and shows the used technologies. 

HMCS-Collab is composed of five modules: 
\begin{itemize}
\item Matching Tool (MT): ensures model matching via two sub-modules: (1) Assisted Matching Tool (AMT) that allows  defining correspondences at metamodel level (meta-correspondences) and (2) Refining Tool (RT) which propagates meta-correspondences to models level by generating the Cartesian product of instances of meta-elements involved in a meta-correspondence, then filtering them thanks to the semantics of their relationships.
\item 
Consistency Management Tool (CMT): ensures the consistency of the system in case of partial models evolution via three sub-modules: (1) Change Detection Tool (CDT) that contains a listener of a fixed set of changes, (2) Consistency Checker Tool (CCT) which analyses the performed changes and their impacts on inter-model correspondences and (3) Inconsistencies Resolver Tool (IRT) which contains a repository of inconsistencies resolutions and recommendations.
\item
Collaboration Tool (CollabT): ensures collaboration mechanisms (e.g. communication, group management and group-awareness) through Communication Tool (CommT), Group Management Tool (GMT) and Group Awareness Tool (GAT).
\item
Decision Management Tool (DMT): contains a set of decision policies and the implementation of their aggregation methods. It is divided in two sub-modules: (1) Decision policies Repository Tool (DRT) which implements  decision making policies and (2) Decision Aggregator Tool (DAT) which assesses the evaluation to obtain the collective decisions about proposals. 
This module and CollabT are invoked by both MT and CMT.
\item
Transformation Tool (TT): supports two kinds of transformation: Model to Text (M2T) and Text to Model (T2M).
\end{itemize}

MT and CMT use Eclipse Modeling Framework (EMF\footnote{https://www.eclipse.org/modeling/emf/}) which is a modeling framework and code generation facility for building tools and other applications based on a structured data model. They both essentially use (i) Graphical Modeling Framework (GMF\footnote{https://www.eclipse.org/modeling/gmp/}) for graphic editors generation, (ii) EMFCollab\footnote{http://qgears.com/products/emfcollab/} for allowing multiple users to simultaneously edit a single EMF model, (iii) Connected Data Objects (CDO\footnote{https://www.eclipse.org/cdo/}) for data persistence, and (iv) Xtext\footnote{https://www.eclipse.org/Xtext/} for developping domain-specific languages by using a powerful grammar language.

CollabT and DMT use Eclipse Communication Framework (ECF\footnote{https://www.eclipse.org/ecf/}) which is a set of frameworks for building communications into applications and services. It provides a lightweight, modular, and fully-compliant implementation of the OSGi Remote Services standard. ECF Shared Object provides basic services for creating replicated objects within a distributed container. The presence API provides services for remote file retrieval and peer-to-peer file transfer. XMPP enhances messages exchange and presence information in real time.

HMCS-Collab provides a web application developed with Java and JSF (JavaServer Faces) to facilitate
the collaboration of local coordinators and to avoid them worrying about the technicality of EMF. Various tools have been used, the most representative ones are: Mysql, PrimeFaces, Hibernate.

\textbf{3. Illustrative example}

To illustrate the GDM enactment, we conduct a part of the alignment process on a  Conference Management System (CMS) (only the matching part). The collaborative matching and the CMS were presented in (Bennani et al., 2018). We do not detail either the process nor the CMS models here, but we can say, briefly, that CMS has been designed from three points of view, leading to three heterogeneous models :

\begin{itemize}
\item 
Software Design (SD) model: represents classes, their attributes and methods.
\item 
Business Process (BP) model: describes roles, activities and products.
\item 
Persistence (PS) model: describes a relational database with tables for data storage.
\end{itemize}

Concerning the collaborative matching process, it works at two levels. First, the actors define correspondences among elements of metamodels called meta-correspondences (MCs). Afterwards, MCs are propagated at models' level and give rise to correspondences. The matching tool MT of HMCS-Collab ensures this propagation, using the semantics of the relationship between elements defined in the metamodel of correspondences (MMC). 

For this application, three Phd students from ADMIR laboratory were involved. Two of them have solid knowledge in model driven engineering. Each PhD student took the role of a local coordinator of a viewpoint model. We call these actors $SD_{LC}$, $BP_{LC}$ and $PS_{LC}$ respectively for SD model, BP model and PS model. The moderator role was played by a senior designer of the team. 
For the CMS example, a proposal consists of a meta-correspondence. Using MMC and the viewpoints' metamodels, each local coordinator specifies the meta-element(s) involved in the meta-correspondence (i.e. meta-elements from his metamodel and the other ones) and the relationships which link them. The body of a proposal contains the meta-elements and the relationship used to relate them. It is expressed according to the following notations:

\begin{itemize}
    \item \textit{Relationship [Metamodel:metaElement $\leftrightarrow$ Metamodel:metaElement]} in case the relationship is symmetric.
     \item \textit{Relationship [Metamodel:metaElement $\rightarrow$ Metamodel:metaElement] }in case the relationship is asymmetric.

\end{itemize}

\begin{figure*}[htbp!]
\center
{\includegraphics[scale=0.45]{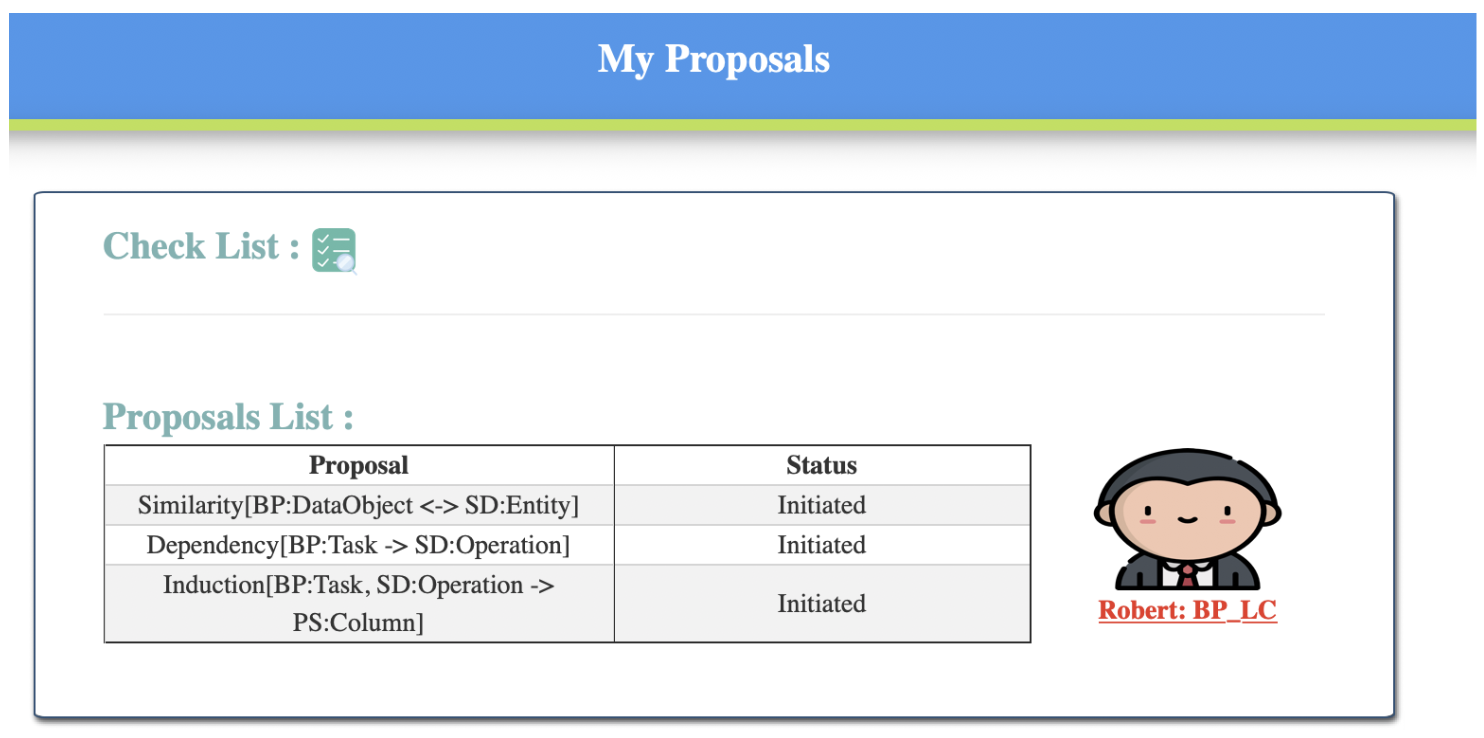}}

\caption{\label{proposals} 
Proposals initiated by Robert, the local coordinator of Business Process model
}
\end{figure*}

Figure \ref{proposals} presents the proposals of the $BP_{LC}$. The first proposal relates the concept \emph{DataObject} of \emph{BP metamodel} with the concept \emph{Entity}  of \emph{SD metamodel} by a \emph{Similarity} relationship. This meta-correspondence means that these two meta-elements may have similar meaning. 
\emph{Dependency} means that a concept depends on another  whereas \emph{Induction} is a special case of dependency where one concept implicates another. 

\begin{figure*}[htbp!]
\center
{\includegraphics[scale=0.45]{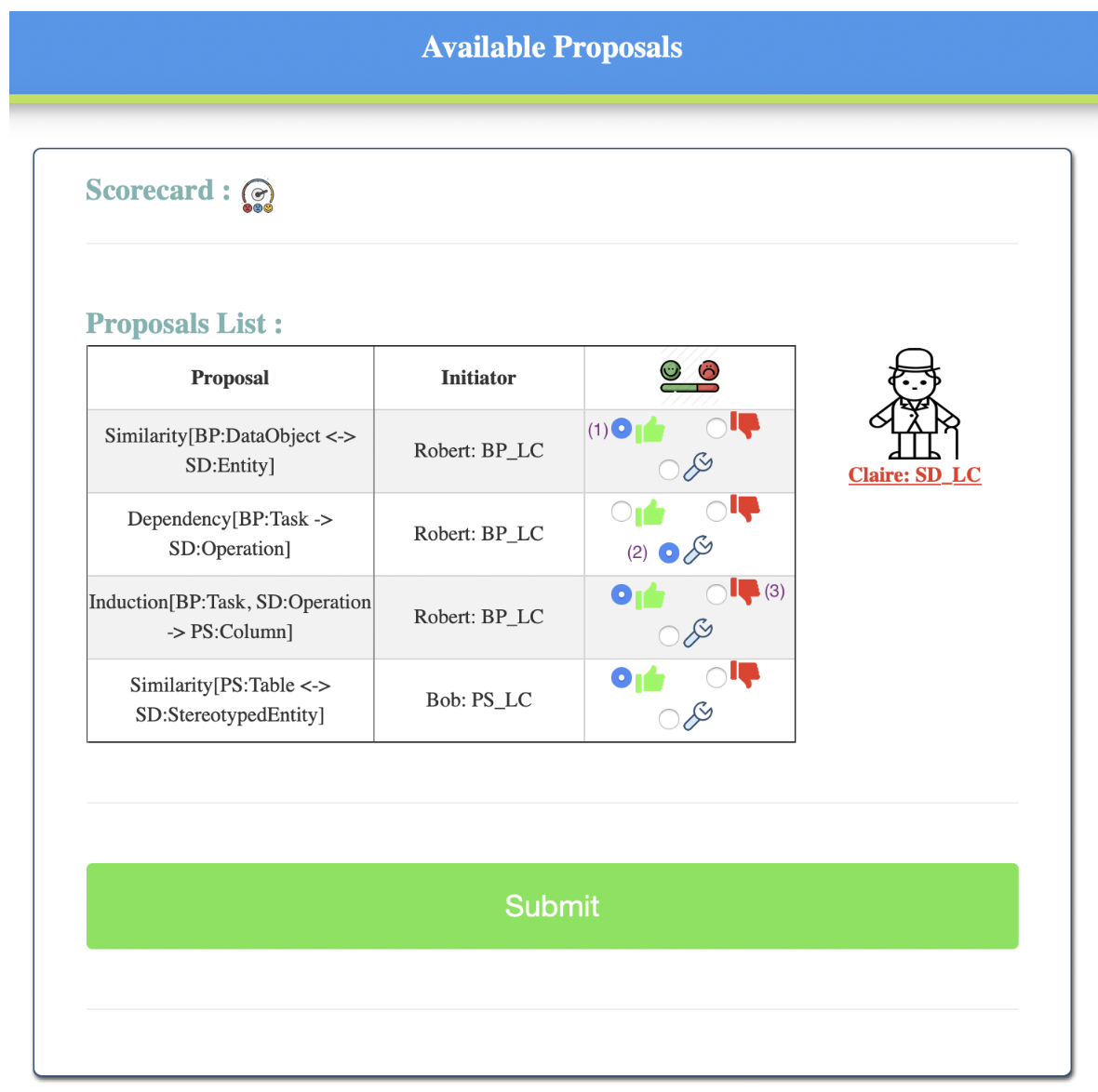}}

\caption{\label{evaluation} 
Proposals having Claire ($SD_{LC}$) as one of the decision makers}
\end{figure*}

After the proposals (MCs) elaboration, they undergo a collaborative evaluation according to the Majority deciding policy (chosen by the moderator). 
Figure \ref{evaluation} summarises the proposals having $SD_{LC}$ (Claire) as a decision maker. In this IHM, $SD_{LC}$ can evaluate each proposal. He or she can either (1) approve, (2) refine or (3) reject each proposal. When she chooses to refine a proposal, she should provide an alternativeProposal and specify if it is conflictual or not with the proposal to which it is associated. This is the case for the second proposal of Figure \ref{evaluation}. So, she has to fill in another IHM (not presented) the description of the alternativeProposal, she specifies it as Induction[BP:Task $\rightarrow$ SD:Operation] and considers it as conflictual with the second proposal of Figure \ref{evaluation}.

\begin{figure*}[htbp!]
\center
{\includegraphics[scale=0.45]{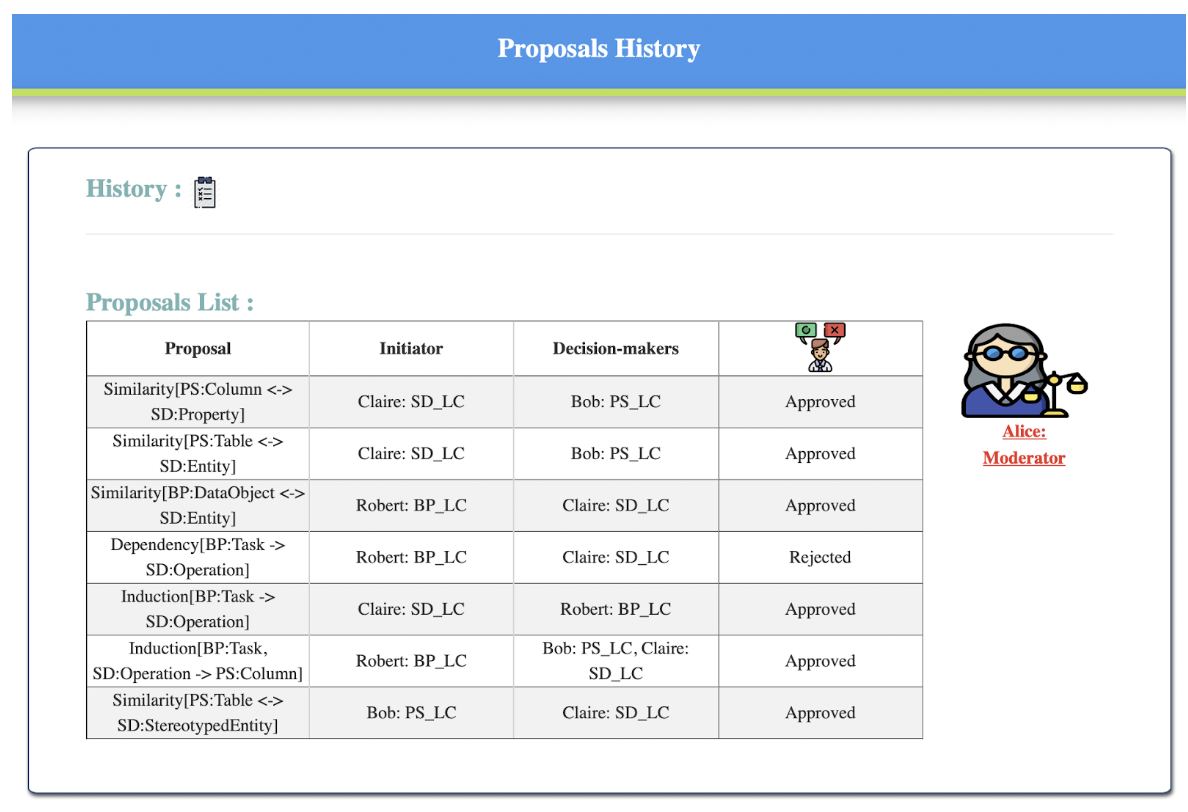}}

\caption{\label{summaryProposals} 
CMS meta-correspondences evaluation summary}
\end{figure*}

Figure \ref{summaryProposals}  summarizes the status of CMS meta-correspondences after the evaluation step. The last column specifies the collectiveDecision for each proposal. All proposals have been approved except Dependency[BP:Task $\rightarrow$ SD:Operation]. Since it was refined and is conflictual with the alternativeProposal, only one of them can be approved.

HMCS-Collab generates correspondences using as inputs (i) viewpoint models; (ii) the set of validated meta-correspondences (i.e. evaluated positively) and (iii) MMC. These correspondences are generated automatically from the meta-orrespondences by exploiting the semantics of their relationships through a \emph{Propagation} process, i.e. a cartesian product of instances of meta-elements involved in an HLC (duplication) followed by a filtering according to the semantics of the relationship used in each MC.  We do not detail the model of correspondences here since it is out of scope of the paper.

\textbf{4. Related work}

We limit our litterature review to approaches describing GDM knowledge since it is the main purpose of this paper.

\emph{4.1 GDM fundamentals}

Group Decision Making (GDM) consists of collaborative activities that aim to develop a collective decision (also called group decision). A GDM process followes usually five steps as defined in (Belton \& Pictet, 1997): 

(i) Define the problem and objective of GDM.

(ii) Identify problem parameters, for instance, \textit{proposals},
and \textit{selection criteria}. 
We call proposal, each solution considered by the actors to meet the objectives set for GDM.  Proposals may be mutually exclusive in case of alternatives or complementary if non-conflictual.
A criterion can be any type of information that makes it possible to evaluate the proposals and compare them. Criteria must be useful, independent and reliable for the given problem. There are many types of criteria: intrinsic characteristics of artifacts or processes, opinions of stakeholders, potential consequences of proposals, etc. This corresponds to qualitative criteria. A quantitative criterion is based on a utility function associating each alternative with a number indicating its expediency according to its consequences. 

(iii) Establish evaluations, i.e. estimate alternatives according to all criteria.

(iv) Select decision making method. i.e., the method by which the group decision will be performed. We can classify existing methods into structured versus unstructured ones (Malavolta et al., 2014). Structured methods are comparison methods based on quantitative criteria (e.g., Analytic Hierarchy Process (AHP), Multi-Attribute Utility Theory (MAUT)), while unstructured methods are qualitative-based comparison methods (e.g., Brainstorming, ranking, SWOT).

(v) Aggregate evaluations, i.e., provide a final aggregated evaluation allowing group decision).

\emph{4.2 Criteria for state-of-the-art analysis}

We devote our literature review to approaches for human group decision-making. In other words, we exclude decision-making approaches based on software agents that aim to assist humans in decision-making since we only consider decision-making processes performed by groups of stakeholders during decentralized and multi-view design. 
To analyze the existing approaches, we use some of the criteria of Saaty and Vargas (2006) and Rekha and Muccini (2014). The retained criteria concern the five following aspects: 
\begin{itemize}
    \item \textbf{Proposals elaboration}. it concerns how proposals are defined. A good approach permits two things: identification and evolution. Identification concerns the définition of the set of proposals. Evolution allows the set of proposals to be extended during the GDM process.

\item \textbf{Management of proposals dependencies}. Dependencies between proposals provide a clear vision on proposals and limit the risks of premature decisions. Different relationships may hold between proposals: specialization, conflict, composition, dependency, override, etc (Malavotla et al., 2014). In this analysis, we consider only two of them: specialization and conflict. The first one helps to express the evolution of proposals during the GDM process, and the second emphasises the incompatibilities between proposals.
\item \textbf{Selection between proposals}. The heart of any GDM process is the indication of  decision makers preferences. This aspect specifies the selection criteria supported by the approach. They can be qualitative criteria versus measurable criteria (quantifiable and previously defined). The criteria can also be weighted according to their importance, in case of qualitative criteria, the weight is associated to decision makers.
\item \textbf{Method of group decision}. it concerns how preferences of different decision makers are taken into account. In particular, we evaluate the adaptability of the methods. A decision method is adaptable if the approach allows its customization (adjusting the acceptance thresholds for example).
\item \textbf{Supporting tool}. This aspect concerns the existence of a tool and specifies the functionalities of the GDM process it supports.
\end{itemize}

\emph{4.3 State-of-the-art synthetic analysis}

The state of the art study focuses on five approaches, namely: Collaboro (Izquierdo \& Cabot, 2016), OntoGDSS (Chai \& Liu, 2010), DSO (Rockwell et al., 2009), MADISE (Kornyshova \& Deneck\`ere, 2010) and Malavolta metamodel (Malavotla et al., 2014). Table \ref{tab3} below summarizes the comparison according to the criteria listed in section 4.2.

\begin{table}[htbp!]
\centering
\caption{\label{tab3}Comparison of GDM modeling approaches}
\begin{tabularx}{\textwidth}{c c c c c c c c}
\toprule
\multicolumn{2}{c}{\textbf{Criteria \textbackslash Approach}} & \textbf{Collaboro}  & \textbf{OntoGDSS}  &
\textbf{DSO} & \textbf{MADISE} & \textbf{Malavolta} & \textbf{MMCollab*}\\
 \midrule
\textbf{Elaboration} & Identification  & + & + & + & + & + & +\\
 & Evolution & + & - & - & + & $\sim$ & +\\
 \midrule
\textbf{Dependencies}  & Specialization & - & - & - & - & + & + \\
& Conflict & + & + & + & + & + & + \\
\midrule
\textbf{Selection criteria}  & Measurable & - & - & + & + & $\sim$ & -\\
& Weighted & - & - & + & + & +  & +\\
\midrule
\textbf{GDM Method} & Adaptable & $\sim$ & - & - &  + &  - & +\\
\midrule
\textbf{Tool} &  & + & - & - & + & $\sim$ & +\\
\bottomrule
\end{tabularx}
\newline
\small{+: criterion at the center of interest of the approach,   $\sim$: criterion partially considered,  
-: criterion not considered at all
\newline *: our metamodel
}
\end{table}

\textbf{Proposals elaboration (Elaboration).}
The five approaches include proposals identification which favors stakeholders involvement unlike when they receive a predetermined list of proposals. For proposals evolution, two approaches out of five: Collaboro and MADISE include the evolution feature during the GDM process. This is possible in the first approach since all of its concepts (proposal, solution and comment) are subject to votes, whereas in the second approach, the attribute nature of alternative concept specifies if  the alternative has evolved. 

\textbf{Management of proposals dependencies (Dependencies).}
All the studied approaches trace conflicts between proposals, either because they treat only alternatives (Rockwell et al., 2009; Chai \& Liu, 2010; Kornyshova \& Deneck\`ere, 2010)  or because they include specific concepts for that (Malavotla et al., 2014; Izquierdo \& Cabot, 2016). Regarding the specialization relationship, it is not traced in most approaches. Only the metamodel of Malavolta  includes this type of link. 

\textbf{Selection between proposals (Selection criteria).}
In Collaboro, alternatives are evaluated through stakeholders votes in order to reach consensus. Decision makers evaluate proposals based on intrinsic criteria (preferences, personal opinions). MADISE lists a set of criteria by subdividing them into subjective criteria (such as intrinsic criteria for decision makers) and quantitative criteria.
The other approaches do not detail enough how the alternatives are evaluated. DMO (from the MADISE approach) and DSO permit to assign a weight to each criterion (quantitative or qualitative), while the meta-model of Malavolta allows to assign a weight only to decision makers.

\textbf{Method of group decision (GDM method).}
DSO and Malavolta's metamodel have been designed independently of any decision-making method. They define a generic concept for the aggregation method but do not propose exploitable methods. Thus, group decision-making processes can not be enacted with these approaches as they are.
Collaboro and OntoGDSS favor a consensual decision-making method developed following the stakeholders' votes. 
MADISE offers a list of aggregation methods that can be used in GDM processes.

\textbf{Supporting tool (Tool).}
OntoGDSS, DSO do not provide any support tool for the decision-making process since their main interest is the construction of the taxonomy needed in group decision-making processes. The tool proposed by Malavolta's metamodel does not integrate in itself the decision methods. The tool proposed by Collaboro supports both proposals elaboration, individual preferences expression and aggregation thanks to its decision engine. MADISE provides a repository of decision-making methods which completes DMO ontology and allows to carry out decision-making situations.

\textbf{5. Conclusion}

In this paper, we first describe the conceptual formalization of group decision-making we propose.  The main advantage of our approach is the description of collaborative decision making concepts and the definition of \textit{decision policies} that are supported by a tool and customizable according to application contexts. Then, we define a practical approach to implement the  decision policies using two well known design patterns (State and Observer). We have investigated the GDM field for a model alignment purpose but the proposed metamodel and concepts can be applicable to other group decision-making situations. 

For future work, we intend to develop a recommendation system to help choosing a decision policy once the proposals and the collaboration characteristics are set. We also aim to enact  MMCollab's concepts in other collaborative contexts.

\textbf{References}

{\leftskip=2em\parindent=-2em

Aksit, M. (1996).Separation and composition of concerns in the object-oriented model. \emph{ACM Comput. Surv.}, 28(4es), 148. \url{https://doi.org/10.1145/242224.242413}

Belton, V., \& Pictet, J. (1997). A framework for group decision using a MCDA model: sharing, aggregating or comparing individual information?. \emph{Journal of decision systems}, 6(3), 283-303.
\url{https://doi.org/10.1080/12460125.1997.10511726}

Bennani, S., El Hamlaoui, M., Nassar, M., Ebersold, S., \& Coulette, B. (2018). Collaborative model-based matching of heterogeneous models. In \emph{2018 IEEE 22nd International Conference on Computer Supported Cooperative Work in Design ((CSCWD))} (pp. 443-448). IEEE.

Bennani, S., Ebersold, S., El Hamlaoui, M., Coulette, B., \& Nassar, M. (2019). A Collaborative Decision Approach for Alignment of Heterogeneous Models. In \emph{2019 IEEE 28th International Conference on Enabling Technologies: Infrastructure for Collaborative Enterprises (WETICE)} (pp. 112-117). IEEE.

Br{\"a}uer, M., \& Lochmann, H. (2008).  An ontology for software models and its practical implications for semantic web reasoning. In: Beckhofer, S., Hauswirth, M.,
Hoffmann, J., Koubarakis, M. (eds.), \emph{The Semantic Web: Research and
Applications} (pp 34-48). Lecture Notes in Computer Science, Vol. 5021. Springer-Verlag,
Berlin Heidelberg New York.
\url{https://doi.org/10.1007/978-3-540-68234-9_6}

Bruneliere, H., Perez, J. G., Wimmer, M., \& Cabot, J. (2015). Emf views: A view mechanism for integrating heterogeneous models. In \emph{International Conference on Conceptual Modeling} (pp. 317-325). Springer, Cham.
\url{https://doi.org/10.1007/978-3-319-25264-3_23}

Chai, J., \& Liu, J. N. (2010). An ontology-driven framework for supporting complex decision process. In \emph{2010 World Automation Congress} (pp. 1-6). IEEE.

Cicchetti, A., Ciccozzi, F., \& Pierantonio, A. (2019). Multi-view approaches for software and system modelling: a systematic literature review. \emph{Software \& Systems Modeling}, 1-27. \url{https://doi.org/10.1007/s10270-018-00713-w}

El Hamlaoui, M., Bennani, S., Nassar, M., Ebersold, S., \& Coulette, B. (2018). A MDE Approach for Heterogeneous Models Consistency. In \emph{Proceedings of the 13th International Conference on Evaluation of Novel Approaches to Software Engineering (ENASE)} (pp. 180-191).
\url{https://doi.org/10.5220/0006774101800191}

Feldmann, S., Kernschmidt, K., Wimmer, M., \& Vogel-Heuser, B. (2019). Managing inter-model inconsistencies in model-based systems engineering: Application in automated production systems engineering. \emph{Journal of Systems and Software, 153}, 105-134. \url{https://doi.org/10.1016/j.jss.2019.03.060}

France, R., Georg, G., \& Ray, I. (2003). Supporting multi-dimensional separation of design concerns. \emph{Proceedings of the Third International Workshop on Aspect-Oriented Modeling with
UML (held with AOSD 2003)}. Boston, Massachusetts, USA. 

Gamma, E. (1995). \emph{Design patterns: elements of reusable object-oriented software}. Pearson Education India.

Golra, F. R., Beugnard, A., Dagnat, F., Guerin, S., \& Guychard, C. (2016). Addressing modularity for heterogeneous multi-model systems using model federation. In \emph{Companion Proceedings of the 15th International Conference on Modularity} (pp. 206-211). ACM. \url{https://doi.org/10.1145/2892664.2892701}

Izquierdo, J. L. C., \& Cabot, J. (2016). Collaboro: a collaborative (meta) modeling tool. \emph{PeerJ Computer Science}, 2, e84.

Kornyshova, E., \& Deneck\`ere, R. (2010). Decision-making ontology for information system engineering. In \emph{International Conference on Conceptual Modeling} (pp. 104-117). Springer, Berlin, Heidelberg.
\url{https://doi.org/10.1007/978-3-642-16373-9_8}

Malavolta, I., Muccini, H., \& Rekha, S. (2014). Enhancing architecture design decisions evolution with group decision making principles. In \emph{International Workshop on Software Engineering for Resilient Systems} (pp. 9-23). Springer, Cham. \url{https://doi.org/10.1007/978-3-319-12241-0_2}

Nuseibeh, B., Easterbrook, S., \& Russo, A. (2000). Leveraging inconsistency in software development. \emph{IEEE Computer 33}, 4, 24–29. \url{https://doi.org/10.1109/2.839317}

Rekha, S., \& Muccini, H. (2014). Suitability of software architecture decision making methods for group decisions. In \emph{European Conference on Software Architecture} (pp. 17-32). Springer, Cham. \url{https://doi.org/10.1007/978-3-319-09970-5_2}

Rockwell, J., Grosse, I. R., Krishnamurty, S., \& Wileden, J. C. (2009). A Decision Support Ontology for collaborative decision making in engineering design. In \emph{2009 International Symposium on Collaborative Technologies and Systems} (pp. 1-9). IEEE. \url{https://doi.org/10.1109/CTS.2009.5067456}

Saaty, T. L., \& Vargas, L. G. (2006). \emph{Decision making with the analytic network process: Economic, Political, Social and Technological Applications with Benefits, Opportunities,
Costs and Risks}. 282. Springer Science+ Business Media, LLC.

Shosha, R., Debruyne, C., \& O'Sullivan, D. (2015). Towards an adaptive tool and method for collaborative ontology mapping. In: Ciuciu I. et al. (Eds.), \emph{On the Move to Meaningful Internet
Systems: OTM 2015 Workshops}. Lecture Notes in Computer Science, vol 9416. Springer, Cham. \url{https://doi.org/10.1007/978-3-319-26138-6_35}

}%

\vspace{0.5cm}

\theendnotes

\vspace{0.5cm}

%\textbf{Copyrights}

%Copyright for this article is retained by the author(s), with first publication rights granted to the journal.

%This is an open-access article distributed under the terms and conditions of the Creative Commons Attribution license (http://creativecommons.org/licenses/by/3.0/).

\end{document}